\documentclass[journal]{IEEEtran}
\usepackage{graphicx}
\usepackage{cite}
\usepackage{amsmath}
\usepackage{hyperref}
\usepackage{atbegshi}

\hypersetup{pdflinkmargin=3pt}
\AtBeginDocument{\AtBeginShipoutNext{\AtBeginShipoutDiscard}}
\makeatletter
\begin{document}

\title{Vehicular Safety Applications and Approaches: A Technical Survey}
\author{Hazem Fahmy, Sabita Mahrajan}

\thanks{}
\thanks{Manuscript received XXX, XX, 2019; revised XXX, XX, 2019.}

\markboth{~Vol.~XX, No.~XX, XXX~2019}{}
\maketitle
\begin{abstract}
This paper proposes an extensive overview of safety applications and approaches as it relates to automated driving from the prospectives of sensor configurations, vehicle dynamics modelling, tyre modeling, and estimation approaches. First, different Advanced-Driver Assistance Systems (ADAS) are introduced along with the main sensing components and technologies. Then, different kinematics modelling of vehicles and tyres are discussed. Finally, various communicational technologies and architectures along with self-driving modules are presented. Moreover, some interesting perspectives for future research are listed based on the extensive experience of the authors.
The objective of this study is to teach and guide the beginner and expert for choosing the most suitable approach for autonomous driving applications in safety and stability targeted issues. 
\end{abstract}


\IEEEpeerreviewmaketitle

\section{Advanced Driving-Assistance Systems (ADAS)}
Advanced Driver Assistance Systems (ADAS) cannot completely prevent accidents. However, they could better protect us from some of the human factors, and human error is the cause of most traffic accidents. The sole purpose of ADAS is to support the driver rather than replace them. To counteract traffic accidents, we could change human behavior and adopt vehicle-related measures and physical road infrastructure-related measures. Another approach is to transition from passive safety measures to active safety measures. 

Passive safety measures include airbags, the structure of the car body, seatbelts, and head restraints. Active safety measures include electronic stability control (ESC), anti-lock braking systems (ABS), and other Advanced Driver Assistance Systems like intersection collision avoidance (ICA) and lane-keeping assistant (LKA). To avoid accidents, we can use ICT (Information and Communication Technology) based ADAS applications, which continuously assist drivers with their driving tasks. Despite the construction difference between various ADAS solutions of many automotive suppliers, the mounting and application of ADAS are similar in the main working principle. Due to the usage of the equipment from Continental in the AutoUniMo project \cite{noauthor_automotive_nodate}, a significant review was prepared based on examples of their ADAS products~\cite{noauthor_advanced_nodate}.

\subsection{Antilock-Braking Systems (ABS)} 

ABS is an electronically regulated scheme that lets drivers retain car control during urgent braking while stopping tires from locking up. Furthermore, by maintaining the tire pressure just below the stage of locking a wheel, ABS guarantees that the highest driving energy is used to control the car and the minimum parking distance is reached.

\subsection{Traction Control Systems (TCS)} 

All driving wheels have some degree of spin, resulting in a mildly more incredible speed than the car itself. When highway circumstances result in too small friction values, the steering wheel can readily start spinning uncontrolled. If the tires start spinning excessively, the car will lose both tractive and side stability. The traction control scheme continually tracks the moving wheels and prevents swinging by decreasing friction energy or pushing the wheel much quicker than the vehicle~\cite{gerstenmeier_traction_1986, demel_abs_1989}. Primarily, anti-spin devices are intended to maintain the vehicle's side stability but also have beneficial friction impacts. 

\subsection{Electronic Stability Control (ESC) or Electronic Stability Program (ESP)} 

These devices initially come from ABS and traction control technologies created for sophisticated engine control~\cite{van_zanten_vdc_1995}. Today, these devices give holistic control of each vehicle wheel. The scheme defines if the vehicle will lose directional control by comparing the engine input with the vehicle's yaw movement and lateral speed. To prevent loss of control and give adjustment of control wheels that are about to spin or slip, precise braking or motor torque can be separately regulated.  

\subsection{Forward Collision Warning (FCW)} 

FCW or Collision Avoidance (CA) is an automotive safety system that decreases crash severity and seriousness. They are also considered pre-crash, crash-mitigating, or forward collision alert systems. It utilizes radar and sometimes laser and camera (both detectors are useless during poor weather conditions) to identify an imminent collision. Once the detection is performed, they either signal the driver when a crash is imminent or take action autonomously without driver feedback (by pressing or pushing, or both). 

Braking collision prevention is suitable at low car velocity (for instance, below 50 km/hour), while driving collision avoidance at greater car velocity. Cars with crash prevention may also have Adaptive Cruise Control and use the same forward-looking detectors. Some of the cars with collision avoidance features available: \cite{Hazem_12c} Audi: A8, A7 (from 2010); A6 (from 2011); A3 (from 2013) Dodge Durango (from 2011) Honda: Accord, Inspire (from 2003) Lexus: LS (from 2003); GS (from 2005); IS (from 2005); RX (from 2008) Skoda Octavia (from 2013) Tesla Model S (2015 model year).

\subsection{Parking assistance (PA)} 

PA is mainly an independent car maneuvering system moving a vehicle from a traffic area to perform parallel, perpendicular or angular parking. It seeks to improve driving convenience and security in local settings, which is accomplished by controlling steering angle and velocity, considering the actual environment condition to guarantee collision-free movement. This concept is to schedule and parameterize the fundamental steering angle and velocity control profile to accomplish the vehicle route within an accessible room. A series of monitored movements are conducted using sensor information and environmental distance measurements. Front and back sensors act as transmitters and receivers, using cameras and RADAR to detect obstacles. Steering and speed checks are calculated and performed in real time.

\subsection{Adaptive Cruise Control (ACC)} 

Autonomous or Adaptive Cruise Control adjusts car velocity to keep a safe distance from ahead cars. Depending on circumstances, the effect is on driver security and highways ' economizing ability. Control relies only on onboard sensor data. ACC utilizes RADAR or LIDAR detectors to decrease velocity and, if needed, back to a previously set velocity. It is regarded as the central part of any potential Intelligent Car generation. 

There are two kinds of ACC: laser-based, which does not detect or monitor unfavorable climate circumstances, and radar-based, which have a pre-crash scheme that informs the rider and offers brake assistance if there is a high probability of crash risk. An expanded variant of ACC is CACC, which stands for Cooperative ACC, which utilizes vehicle-to-vehicle data in the same lane as feedback. It also controls the distance to the front vehicle, even in stop-and-go situations. It warns a rider or deliberately slows velocity if the relative distance range becomes too low. This implementation supports the driver, especially in congested traffic and tailbacks. Moreover, safety is improved by the pre-defined range and alert when urgent braking is needed \cite{vollrath_ableitung_2006}. 


\subsection{Emergency Braking Systems (EBS)} 

EBS or Automatic Emergency Brake (AEB) or Emergency Brake Assist (EBA) are safety systems which identify critical traffic conditions and guarantee optimum braking. The steering scheme is on standby alert when EBA detects an early crash. Next, the rider is warned, and a slight pre-brake starts to save precious stop distance. EBA's owned smart cities could deter many low-speed rear-end accidents, incredibly optimized for metropolitan vehicle use. It also enhances riding safety through effective braking and braking in hazardous circumstances. Therefore, rear-end crashes can be prevented entirely. Furthermore, the effects of crashes are decreased by reducing effect velocity and post-impact energy effect. Emergency brake aid is also a feasible interface for pre-crash and restraint devices or pedestrian protection \cite{vollrath_ableitung_2006}.

\subsection{Lane Change Assist (LCA)} 

LCA or or Lane Departure Warning (LDW) or Lane Keeping (LK) depend on a lane algorithm estimation data. One significant stage in the route assessment process is to extract readings or detections that can be used to predict highway or lane shapes. White lane markers or the highway boundary itself generate these detections. For many years, lane assessment was under severe growth using a gray-scale camera. 

Under certain conditions, passive camera-based devices can be degraded, e.g., by dramatic modifications in ambient light. The system informs the driver that a visual or haptic alert, such as steering wheel vibrations, accidentally switched lanes. It also checks the roadside and detects when a car leaves the lane or highway. The scheme can assess whether the lane shift is deliberate by regulating steering motion. Traffic crashes induced by road-leaving vehicles or collisions with passing or parked cars can be reduced \cite{noauthor_advanced_nodate}.

\subsection{Blind Spot Detection (BSD)} 

Blind Spot Information System (BLIS) is Volvo's protective system. As the name indicates, this instrument identifies cars, barriers, driver-side, and rear individuals. Sometimes it also involves Cross-Traffic Alert alerts riders from a parking room when traffic approaches from the sides. George Platzer obtained the Blind Spot Monitor's patent, which is an elegant, and comparatively cheap option. Using two door-mounted glasses, it checks the blind spot region for an impending crash. When a vehicle reaches the blind place while a rider switches paths, it generates a noticeable warning. BSD controls the vehicle-side region and offers a feature that alerts a user with a visual, such as a sign in the side-view mirror \cite{noauthor_advanced_nodate} or an audible signal when objects are in the blind areas. This scheme aims to prevent prospective crashes, particularly during heavy traffic lane shift maneuvers \cite{vollrath_ableitung_2006}.

\subsection{Rear Cross Traffic Alert (RCTA)} 

RCTA may allow accidents to be avoided when reversing from car parks that frequently involve personal injuries. The solution is based on two short-range radar sensors monitoring an angle of 120 degrees. When an impending crash is detected, an alert is given to the driver through sound and lights LEDs in the interior mirror. ADAS can break the vehicle automatically if the alert sounds. 

This system can be used to reliably calculate the collision trajectory and speed of the crossing vehicle and the distance that this new warning function can improve the ADAS. RCTA can also help avoid accidents when a driver leaves a car park, often leading to severe injuries to pedestrians or cyclists. The environment behind a vehicle is monitored and checked for objects for this function. When an object is detected in the opposite direction, the driver receives both audible and visible warnings \cite{noauthor_advanced_nodate}.

\subsection{Intelligent Headlamp Control (IHC)} 

Adaptive High Beam Assist is a marketing name for Mercedes-Benz concerning a specific headlight control strategy. It automatically adjusts the headlamp range continuously so that the beam reaches the next vehicle. This technique ensures maximum visibility without disturbing other road users. 

The Mercedes E-Class was launched for the first time in 2009. It enhances the range with a continuous beam range, from a low-targeted beam to a high-targeted beam, where the beam range can vary from 65 to 300 meters. IHC also automatically regulates the lights of a vehicle following environmental conditions. This application optimizes changes in nighttime drives between the entire beam and the dipped headlights. Therefore, driving through tunnels at night is comfortable and safer. In addition, drivers in incoming cars are no longer blinded to the lights of a vehicle \cite{noauthor_advanced_nodate}.

\subsection{Driver Drowsiness Detection (DDD)} 

DDD is a car safety technology that helps prevent driver drowsiness accidents. Several studies have shown that around 20\% of all road accidents are related to fatigue. Some existing systems learn driver patterns and can detect when the driver is drowning. 

The different technologies used to detect driver drowsiness: steering pattern surveillance, vehicle position in lane surveillance, monitoring of the eye and face of the driver, and psychological measurement.

\section{Sensing Technologies}

\begin{figure*}
\label{fig:adas}
  \centering
  \includegraphics[width=\textwidth]{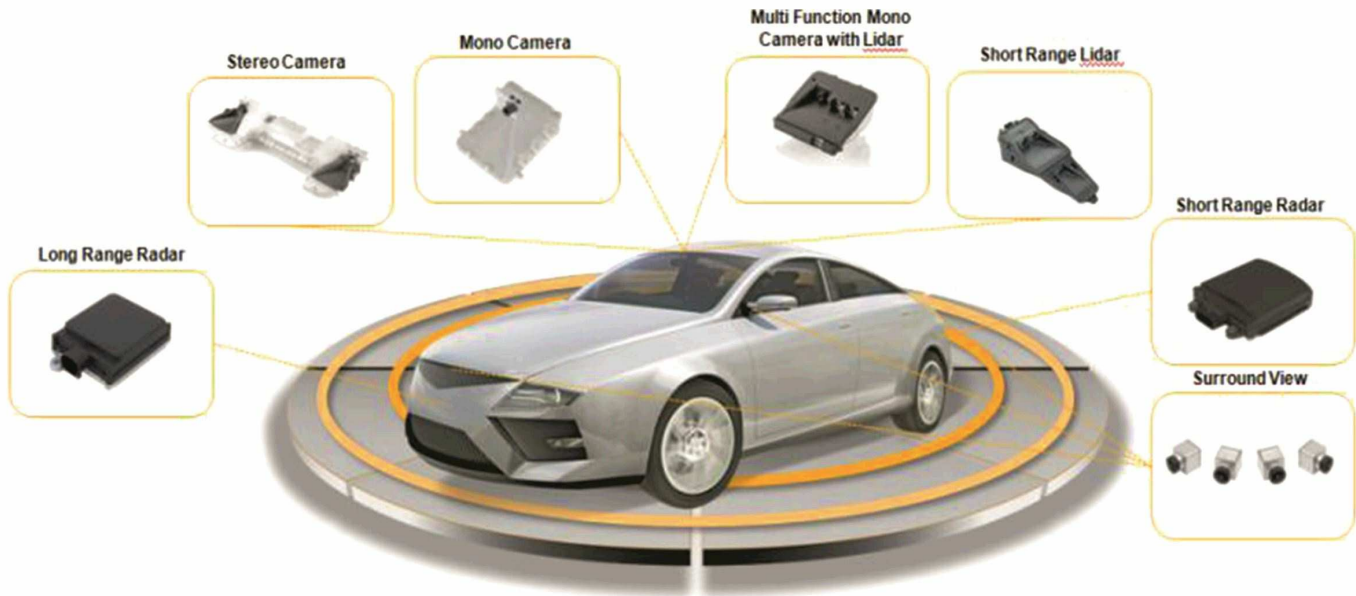}
  \caption{An example of the ADAS main components~\cite{noauthor_advanced_nodate}.}
\end{figure*}


\subsection{Internal Sensory Systems}
Many of the outputs used for the independent car devices are based on existing sensors and schemes. These signals are easily accessible on the CAN of an internal vehicle's systems and are used to provide vehicle data and are also used to produce independent sensor output reactions. Because many of these devices and parts are not usually intended for extra cargo independent devices, many of these detectors or devices involve extra robustness to safeguard against increased loads. Some of these signals and interacting systems include \cite{Fleming_2}:

\begin{itemize}
\item Wheel Speed Sensor is generally a Hall Effect sensor that is situated on each wheel and produces a signal that is then converted into the wheel velocity. This can be interpreted as a tachometer that is mainly used by ABS and contemporary cruise control systems.

\item Yaw Rate Sensor is usually a piezoelectric tool that uses the Coriolis force to define the curve or the dimensional differences between the orientation of the car and the current direction of motion. This is essential in order to regulate electronic stabilization.

\item Lateral / Longitudinal Sensors are MEMS or piezoelectric sensors that are used for lateral / longitudinal speed and for the digital stability command in contemporary car for collision detection and electronic stability control.

\item Steering Inputs in autonomous cars are obtained using digital energy steering signals and capacities to comprehend and request different responses. The particular variables controlled include the speed sensor and location of the steering wheel.

\item Driver Inputs in modern cars include brake requests, accelerator pedal, signal turning, steering power, headlights, window wipers and equipment choice.
\end{itemize}

\subsection{External Sensory Systems}
\subsubsection{Ultrasonic Sensors} Vehicle reversing and parking aids commonly use low-cost short-range ultrasonic sensors. The detectors work at 50 kHz neighbourhood frequencies. They are concurrently transmitted and received with a piezoelectric membrane component by short ultrasonic waves. There is not enough beamwidth in a single detector. Consequently, complete coverage of the side-section of a vehicle involves the acquisition of signals from four detectors. Circuitry for signal processing is incorporated into the detector. The detection variety of ultrasonic detectors is about 2.5 m. Development attempts are being made to expand the range to 4.0 m. The detectors are installed on the seat fascia of the car and look like a linear array of four circular dimples \cite{Fleming_100}.

\subsubsection{Radars} These frequency-modulated/continue-wave radar devices directly measures the range and closing speed. The difference between the transmission modes and the doppler-shifted received signal frequencies is calculated. In order to obtain the range of the car, sums of the beat signals and frequencies are created (doppler signals disappear when beat frequencies are combined). Differences between beat frequences, on the other side, indicate the distance closing rate \cite{skolnik_introduction_1983}. This concurrently measures the target range and the closing rate. 

Monopulse radar is another type of radars where the range and azimuthal angle data are acquired from single broadcast pulses covering a broad forward region. Sum and distinctions of observed wavefront stages of reflected pulse are identified with radar receiver antennas from the dual side-by-side \cite{Fleming_89, skolnik_introduction_1983}. The objective range is extracted from the pulse transit time of the total signal, the range closing rate comes through monitoring of the target range data versus time, and the target azimuthal angle is obtained from the phase variation of the wavefront obtained, as identified by the antenna components obtained side by side. When a target is straight ahead, the double antennas detect the received pulse-wave-front simultaneously and the azimuthal phase angle is zero. If the target is situated at the right of the car, the reception antenna will lead to the identified stage of the right-hand antenna and the azimuthal phase angle will be of a positive value and vice versa if the target is on the left. In certain luxury European cars and heavy trucks both in North America and Europe, monopulse radar is heavily utilized.

Furthermore, LIDAR relates to a light identification and ranging device, which delivers millions of light pulses per second in a well-designed model. It can generate a vibrant, three-dimensional map of the setting using its spinning axis. For most of the existing autonomous vehicles, LIDAR is the heart for object detection. In a real scene, the LIDAR points are never ideal. The problems with managing LIDAR points are due to scanner sparsity, lack of points and unorganized patterns. The setting also presents additional difficulties to the understanding, as the surfaces can be arbitrary and uneven. Sometimes even humans find it hard to interpret helpful data from a scan point visualization.

Additionally, night vision sensors look ahead to the highway and captures night-time traffic scene. These systems design a picture of the scene on the heads-up or instrument panel screens to support the driver. For night vision, two distinct techniques are applied: a) far-infrared (FIR) sensors that detect long-wave IR warm body heat radiation; b) near-infrared (NIR) sensors that project short-wavelength non-visible IR illumination that deliver day-like road pictures.

Finally, pulsed doppler radar transmits pulses (bursts) of continous-wave signals, which additionally include a doppler frequency shift when recording a moving destination, to discriminate against shifting out of the fixed targets \cite{skolnik_introduction_1983}. The gallium arsenide MMIC circuits quickly switch three beams that sequentially scan the azimuthal regions on the right, middle and left of the highway \cite{Fleming_87}. The pulse-doppler radar output signal offers a range, a range closing rate and target azimuthal location. Target range is based on pulse travel transit time, while range closing rate is obtained from the doppler frequency shift of transmitted signal, and target azimuthal location is obtained from understanding information of the destination which is identified by one of the three beams or mixture of the beams. In certain luxury European and North American cars Pulsed doppler radar is used.
 
\subsection{Positioning} The combination of satellite navigation technologies and inertial navigation technologies is one of the most common methods to locate a car. The worldwide location of the car can be regularly remedied by satellite navigation technologies, such as GPS and GLONASS. Depending on the signal intensity and performance of the tools used, their precision can differ from a few of tens of meters to some millimeters. External equipment is not required for inertial navigation devices which use accelerometer, gyroscope and signal processing methods to predict car orientation. Without addition of other sensors, however, it can be difficult to initiate an inertial navigation system and the error is unlimited over time. GPS utilization needs credible service transmissions from external satellites in localisation. This technique is only accurate when the GPS signal and dead reckoning odometry of the car are accurate in addition to that costly, and high-precision sensors are required. A few excellent instances of problem fields are localization of indoor environment, subterranean and urban canyon settings, where large structures reject the vehicle's accurate GPS signal measurements. The use of three gyroscopes and three accelerometers is usually called the inertial unit or IMU of the vehicle. This is essential for structures that require a strong level of understanding the movement of cars in typical independent cars using GPS.

\subsection{Camera Vision} Vehicle camera-vision serves either of two applications: i) scene viewing, for example as used for camera backup screens and (ii) machine vision scene comprehension or understanding, for example if nobody is genuinely seeing the video, as is the case in the vehicle lane departure alert. For use in automotive image interpretation apps, vibrant adaptability of the camera vision must be 120 dB (to enable the camera to generate transparent pictures under all lighting circumstances). In a severe setting with varying temperatures, vehicle cells must also operate confidently with longer durability than consumer-oriented cameras \cite{Fleming_99}. Space around the rear-view window where these cameras are installed is important, so the frame size is also another important variable. Although the camera view has excellent lateral visibility (i.e. excellent image volume size), its scope and measurement accuracy are low. On the other side, the radar is very precise, but its lateral resolution is restricted. As such, radar and computer vision techniques are often coupled to reliably identify object magnitude and variety by using sensor fusion \cite{Fleming_100}.

\section{Modelling} 
Development of advanced vehicle systems require intense effort of modeling, since modeling is the most feasible approach toward breaking-down the physical and mathematical properties in the driving environment. In order to model, objects to be modeled must be classified first. In the road environment, modeling could be done for the driver, vehicle's properties, relation between tires and roads, road geometry, and finally a mastermind that controls all the modeling process. In this section, different vehicle models and dynamics mimicing in addition to tire models and estimation of tire-road relation with prediction/estimation models will be introduced and investigated.

\subsection{Vehicle Dynamics Modelling} 

A vehicle's jerky movement is decomposed into two components: stable motion, unstable motion. The stable movement is implemented primarily by steering command assuming a smooth surface; the unstable movement is due to irregular terrain parts. Because a kinematic system is applicable in modeling fluid movement, a steady translation and rotation system is used. On the other side, a four-wheel car model comparable to designs used for ideal suspension system layout is used to define unstable motion \cite{hady_theoretical_1989}. It represents three unstabilized parts: bounce, pitch, roll. Using the movement equations of the Lagrange, the unstabilized car model movement can be readily discovered. This leads to a stronger depiction of vehicle movement. Subsequently, a picture series obtained from a sensor rigidly connected to the car with recognized orientation is presumed to be accessible to assess the vehicle's movement as well as constructions of a scene. The picture plane displacements of constructions presented by the vehicle's movement are then obtained. 

Studies based on this approach use the system's dynamic model in which some states can be analyzed (like the angular wheel velocity) and some other states can not be measured (like friction force, rotational speed, etc.). Based on dynamic system model and measured states, the remaining states can be predicted using distinct assessment techniques such as recursive least square (RLS), Kalman filter, etc. Some of the most popular dynamic designs are wheel tyre model, spin dynamic model, bike model, quarter-car model, and four-wheel car dynamic model, described in more detail next. 

\subsubsection{Wheel Dynamics Modelling} is used with a tyre model to predict longitudinal force and longitudinal friction \cite{canudas-de-wit_new_2003, tesheng_hsiao_robust_2011}. In \cite{tesheng_hsiao_robust_2011}, the measured wheel force and angular speed in Eq. (3), acquired from the time equilibrium equation of each wheel, was replaced to assess longitudinal tyre force. The work in \cite{rajamani_algorithms_2012} used angular wheel velocity measured information and suggested a sliding mode observer (SMO) to predict longitudinal tyre force based on single wheel dynamic model. Work done by \cite{wanki_cho_estimation_2010} used the same strategy to assess longitudinal tyre strength and proved that the projected strength is precise for maneuvering small slip vehicles. Using a single wheel template with cascaded first and second sliding observers  \cite{Khal_66, msirdi_first_2006} estimated contact force. They used measured car longitudinal information, wheel and wheel torque angular location along with solid differentiators and SMO to predict wheel speed and acceleration, lateral and vertical tyre forces and friction coefficient.

\begin{equation}
\begin{array}{c}{m_{\mathrm{w}} \dot{v}_{x}=F_{x}-F_{\mathrm{rr}}} \\ {J_{\mathrm{w}} \omega_{\mathrm{w}}=\left(T_{\mathrm{w}}-T_{\mathrm{b}}\right)-F_{x} r_{\mathrm{w}}-F_{\mathrm{rr}} r_{\mathrm{w}}}\end{array}
\end{equation}

where $w_{m}$ is the total mass of the wheel, $w_{J}$ is the moment of inertia of the wheel, and $T_{w}$ ,$T_{b}$ are the drive and brake torques, respectively.

\subsubsection{Kinematic model} The most popular car designs used to predict car roll dynamics are 3DOF models depicting car lateral, yaw and roll movements, and 1DOF designs depicting roll dynamics. The 1DOF roll dynamic model has practical advantages compared to the 3DOF model, which doesn't need the cornering stiffness (which isn't simple to estimate), besides not being susceptible to nonlinear tyre dynamics. Therefore, 1DOF design is commonly used in car condition assessment research. The car roll dynamic system is mostly used to predict the car roll angle, a main consideration in obtaining standard wheel load. Assuming the roll axis is fixed and there is no vertical motion, the equation of motion is formulated as: 

\begin{equation}
\begin{split}
(I_{x}+m_{\mathrm{s}} h_{\mathrm{roll}}^{2}) \ddot{\phi}_{\mathrm{chassis}}+c_{\mathrm{rol}} \dot{\phi}_{\mathrm{chassis}}+k_{\mathrm{roll}} \phi_{\mathrm{chassis}}\\=-m_{\mathrm{s}} h_{\mathrm{roll}} a_{y, \mathrm{m}}
\end{split}
\end{equation}

where $I_x$ + $m_sh^2_{roll}$ is the moment of inertia around the roll axis, $I_x$ is the moment of inertia around $x$ axis and m is the sprung mass of the vehicle.

\subsubsection{Quarter Car Model} Quarter Car Model is a 2DOF model, mostly used to model the car's vertical structure (particularly suspension). Quarter vehicle model primarily used in friction assessment research to achieve standard tyre force and highway profile. Several trials evaluated the vertical speed of the un-sprung body and the deflection of the vehicle, using the following equation to evaluate standard force \cite{Khal_75, jin-oh_hahn_gps-based_2002, gustafsson_slip-based_1997, doumiati_estimation_2011, doumiati_road_2012} used the quarter-car template to evaluate normal tyre load and road profile. First, they used accelerometer readings to calculate the vertical stance of the car body, then used it as a measured condition for Kalman filter to assess standard wheel load and traffic profile. Next, the vehicle's planar dynamic models are presented; four-wheel car design and its streamlined variant are addressed in more detail, which are commonly used in car condition assessment research. The equations of the motion for the quarter-car vehicle model 
are presented as:

\begin{equation}
\begin{array}{l}{m_{\mathrm{s}} \ddot{x}_{\mathrm{s}}+c_{\mathrm{s}}\left(\dot{x}_{\mathrm{s}}-\dot{x}_{\mathrm{u}}\right)+k_{\mathrm{s}}\left(x_{\mathrm{s}}-x_{\mathrm{u}}\right)=0} \\ {m_{\mathrm{u}} \ddot{x}_{\mathrm{u}}+c_{\mathrm{s}}\left(\dot{x}_{\mathrm{u}}-\dot{x}_{\mathrm{s}}\right)+\left(k_{\mathrm{u}}+k_{\mathrm{s}}\right) x_{\mathrm{u}}-k_{\mathrm{s}} x_{\mathrm{s}}=0}\end{array}
\end{equation}


\subsubsection{Four-wheel Car Model} Four-wheel car model (also known as two-track model) only sees the vehicle's longitudinal, lateral and yaw movements while ignoring spin, pitch, and vertical movement.

The equations of motion for this model are described 
as follows~\cite{doumiati_vehicle_2013}:
\begin{equation}
\footnotesize
\dot{\beta}=-\dot{\psi}+\frac{1}{m_{\mathrm{v}} V_{\mathrm{g}}} \left[ \begin{array}{c}{-\left(F_{x 11}+F_{x 12}\right) \sin (\beta-\alpha)} \\ {+F_{y 11} \cos (\beta-\alpha)+F_{y 12} \cos (\beta-\alpha)} \\ {+F_{y 11} \cos (\beta-\alpha)+\left(F_{y 21}+F_{y 22}\right) \cos (\beta-\alpha)} \\ {-\left(F_{x 21}+F_{x 22}\right) \sin \beta}\end{array}\right]
\end{equation}

\begin{equation}
a_{y}=\frac{1}{m_{\mathrm{v}}} \left[ \begin{array}{c}{F_{y 11} \cos \delta+F_{y 12} \cos \delta+F_{y 21}+F_{y 22}} \\ {+F_{x 11} \sin \delta+F_{x 12} \sin \delta}\end{array}\right]
\end{equation}

\begin{equation}
a_{x}=\frac{1}{m_{\mathrm{v}}} \left[ \begin{array}{c}{-F_{y 11} \sin \delta-F_{y 12} \sin \delta+F_{x 21}+F_{x 22}} \\ {+F_{x 11} \cos \delta+F_{x 12} \cos \delta}\end{array}\right]
\end{equation}

\begin{equation}
\begin{array}{r}{\dot{V}_{x}=V_{y} \dot{\psi}+a_{x}} \\ {\dot{V}_{y}=-V_{x} \psi+a_{y}}\end{array}
\end{equation}

\begin{equation}
\footnotesize
V_{\mathrm{g}}=\frac{1}{m_{\mathrm{v}}} \left[ \begin{array}{c}{\left(F_{x 11}+F_{x 12}\right) \cos (\beta-\delta)+F_{y 11} \sin (\beta-\delta)} \\ {+F_{y 12} \sin (\beta-\delta)+\left(F_{x 21}+F_{x 22}\right) \cos \beta} \\ {+\left(F_{x 21}+F_{x 22}\right) \cos \beta} \\ {+\left(F_{y 21}+F_{y 2}\right) \sin \beta}\end{array}\right]
\end{equation}

where $I_z$ is the moment of inertia of the car around $z$ axis, $m_v$ is the vehicle mass and $\beta$ is the vehicle side-slip angle.


\subsubsection{Bicycle model} The bike model was first introduced by Segel in 1956 \cite{segel_theoretical_1956}. Bicycle model is commonly used to define the vehicle's operating characteristics, where vertical and roll movements are not considered. Many studies have used this design car along with some prediction algorithms to assess lateral car conditions, friction force and/or coefficient. Tyre-based algorithms are implemented next. The simplified equations of motion for the bicycle
model are as follow:
\begin{equation}
\ddot{\psi}=\frac{1}{I_{z}}\left[l_{f}\left[F_{x 1} \sin \delta+F_{y 1} \cos \delta\right]-l_{\mathrm{r}} F_{y 2}\right]
\end{equation}

\begin{equation}
\dot{\beta}=\frac{1}{m_{\mathrm{v}} V_{\mathrm{g}}} \left[ \begin{array}{c}{-F_{x 1} \sin (\beta-\delta)+F_{y 1} \sin (\beta-\delta)} \\ {+F_{y 2} \cos \beta-F_{x 2} \sin \beta}\end{array}\right\}
\end{equation}

\begin{equation}
\dot{V}_{\mathrm{g}}=\frac{1}{m_{\mathrm{v}}} \left[ \begin{array}{c}{F_{x 1} \cos (\delta-\beta)-F_{y 1} \sin (\delta-\beta)} \\ {+F_{x 2} \cos \beta+F_{y 2} \sin \beta}\end{array}\right]
\end{equation}

\subsection{Tyres and Vehicle-Road Relation Modelling} 







Vehicle-dynamic estimates were commonly debated and investigated in literature \cite{Baffet_1, Baffet_2, gustafsson_slip-based_1997, kiencke_observation_1997, Baffet_5, Baffet_6, ono_estimation_2003, ray_nonlinear_1996, stephant_virtual_2004, Baffet_10, ungoren_study_2004}. Vehicle–road scheme can be built by merging a single-track model \cite{segel_theoretical_1956} with a linear tire–force model. This mixed system, recognized as the linear single-track system, operates well as long as vehicle requirements stay low. However, the car dynamic becomes nonlinear when a car experiences elevated accelerations or when highway friction shifts, so the linear single-track system is no longer adequate. Several surveys have defined participants that consider differences in cornering stiffness (or highway friction) \cite{Baffet_5, ray_nonlinear_1996, msirdi_vehicleroad_2005}. In \cite{Baffet_5} tire-road friction is regarded a disruption, whereas in \cite{msirdi_vehicleroad_2005} tire-road parameters are detected with an observer, while in \cite{ray_nonlinear_1996} tire-road forces are modeled on an embedded random walk system.


Many tire-force models have been proposed, from empirical methods (Pacejka model \cite{pacejka_magic_1992}, Burckhardt model \cite{burckhardt_fahrwerktechnik._1993}, Kiencke model \cite{kiencke_observation_1997}) to analytical methods (LuGre model \cite{canudas-de-wit_dynamic_2003}). Three distinct stages of tire handling behaviours can be distinguished (the linear, transient and saturation stages). In the first stage, lateral tire forces are linear with respect to sideslip angle. The second and third stages, nonlinear, occur when the vehicle approaches the physical limit of adhesion. This study uses the nominal linear model and proposes the linear adaptive model.

An SMO is used to predict the car sideslip angle while using a linear horizontal friction force template and treating longitudinal friction stresses as recognized observer inputs \cite{Cao_13}. The linear connection depicting lateral friction forces, however, is no longer adequately precise when the lateral speed is above 4 $m/s^2$ and tyre sideslip angles are large. Moreover, in a modern advanced vehicle, longitudinal friction forces aren't always accessible. In \cite{grip_nonlinear_2008}, the non-linear state observer for car velocity assessment with assured input-to-state stabilization (ISS) \cite{Cao_14} is provided. In relation to wheel angular velocity and steering wheel angle readings, it is focused on longitudinal and lateral accelerations and yaw frequency estimates. The sideslip value is also achieved by longitudinal and lateral velocities. The observer gains must be determined at each test moment, however. Thus, the assessment algorithm's real-time efficiency may be impacted, and measurement mistakes may interfere with the selection of observer gains.

Tire models designs usually convey the connection between tire forces and moments with slip-ratio / slip-angle used in numerous research to evaluate friction forces and friction coefficient. It is assumed that the tire forces and moment, slip ratio and/or slip angle are available (or can be estimated) and the model parameters and friction coefficient are estimated by comparing force / moment data with different tire models. Different mathematical tyre models have been created depending on the tyre template and the time conduct that can be caught (steady-state or temporary)~\cite{Khal_68}; some of them more popular for friction assessment purposes are presented in this paper. 


\subsubsection{Pacejka tire model (which is also called Magic Formula)} is a semi-empirical tyre system launched by Pacejka in 1992 \cite{pacejka_magic_1992}. The model utilizes unique features to depict longitudinal, lateral and alignment moment forces. The camber angle, cornering stiffness, and load variation are also taken into consideration in the current variant of the magic formula. Magic formula was commonly used in car states literature and tyre friction estimates \cite{pacejka_magic_1992, bakker_tyre_1987, pacejka_shear_1991}. Van Oosten et al. \cite{van_oosten_determination_1992} explained the parameters of Magic Formula from the experimental information and addressed all the problems associated. Kim et al. \cite{Khal_117} used the traditional notion of friction to formulate friction, assuming the Magic Formula numbers are known. Using an instrumented car, they projected forces on each wheel using Magic Formula. They then predicted coefficients of longitudinal and lateral friction. Yi et al. \cite{yi_estimation_1999} used an observer-based method to predict the tyre street friction coefficient using the Magic Formula with recognized parameters for longitudinal forces. First, they used a sliding mode observer (the angular frame speed was assessed) to assess the car states, then they estimated the tyre street friction ratio using a recursive least square algorithm. Jayachandran et al. \cite{jayachandran_fuzzy_2013} created an algorithm based on Fuzzy Logic to assess the importance of longitudinal and lateral stresses, also aligning time depending on slip ratio and slip angle. They used Magic Formula with recognized coefficients to calculate the tire forces and moment based on distinct slip values; then they described the fuzzy-logic algorithm's affiliation feature between the inputs (slip ratio, slip angle) and outputs (tyre angles, aligning moment) based on calculated values. The formulation of this tire model for longitudinal and lateral force and aligning moment are as follows:
\begin{equation}
\begin{array}{l}{F_{x}\left(s+S_{\mathrm{hx}}\right)=} \\ {D_{x} \sin \left[C_{x} \arctan \left(B_{x} s-E_{x}\left(B_{x} s-\arctan \left(B_{x} s\right)\right)\right)\right]+S_{\mathrm{vx}}} \\ {F_{y}\left(\alpha+S_{\mathrm{hy}}\right)=} \\ {D_{y} \sin \left[C_{y} \arctan \left(B_{y} \alpha-E_{y}\left(B_{y} \alpha-\arctan \left(B_{y} \alpha\right)\right)\right)\right]+S_{\mathrm{vy}}} \\ {M_{z}\left(\alpha+S_{\mathrm{hz}}\right)=} \\ {D_{z} \sin \left[C_{z} \arctan \left(B_{z} \alpha-E_{z}\left(B_{z} \alpha-\arctan \left(B_{z} \alpha\right)\right)\right)\right]+S_{\mathrm{vz}}}\end{array}
\end{equation}

\subsubsection{Dugoff Tyre Model} is a physical model Dugoff et al. launched in 1969 \cite{Khal_120}. In the tyre contact patch, a standardized vertical pressure distribution is presumed. In its easiest form, Dugoff framework reflects the relationship between longitudinal and lateral force and drop as a result of two parameters; the tyre stiffness, ($C_x$, $C_y$), which explains the path of force-slip curve in small slip area and the tire-road friction coefficient) (which defines its curvature and maximum value. In the previous sense, some research that used the Dugoff tyre template to assess friction or associated parameters are implemented. Ghandour et al. \cite{Khal_89} created an algorithm using Dugoff to replicate the lateral friction coefficient. First, they created a lateral force and slip angle assessment algorithm. They then measured the friction ratio using nonlinear regression technique to minimize the error between projected lateral force and Dugoff design lateral force. Nilanjan et al. \cite{patra_observer_2012} suggested a sliding mode observer with altered Dugoff model to assess rotational speed and friction coefficient; the only recorded condition was angular speed of the wheel. Doumiati et al. \cite{doumiati_onboard_2011} suggested a real-time method to assess tire-road lateral forces and side-slip angle using two prediction methods of expanded and unscented Kalman filter, in which lateral force was simulated using Dugoff method. Four-wheel car design was used as the system's dynamic model, with longitudinal and lateral speed, yaw and spin speed, left and right suspension deflection, and angular velocity of each wheel. In order to estimate the longitudinal and lateral force and slip ratio, a weighted Dugoff model is used to estimate the friction coefficient, for instance for the longitudinal direction simplifying the Dugoff equation gives:
\begin{equation}
\mu_{x-\max }^{2} F_{z}^{2}-2 \mu_{x-\max }\left|C_{x} \tau\right| F_{z}+\left|C_{x} \tau\right| F_{z}=0
\end{equation}

where in this equation $\tau = \frac{s}{1+s}$. The solution for this equation is expressed as:
\begin{equation}
\mu_{x-\max }=\frac{\left(\left|C_{x} \tau\right| \pm \sqrt{C_{x} \tau\left(C_{x} \tau-F_{x}\right)}\right)}{F_{z}}
\end{equation}

\subsubsection{Brush tyre model} In this model, the ground region in touch with the street can be represented as infinitesimal bristles. As shown in Khal-Fig14, Contact patch region divided into two regions \cite{svendenius_experimental_2009}: adhesion and slipping/sliding. The bristles pass the force through mechanical adhesion in the first area, and the slide of the bristles on the highway in the second area outcomes in friction force resistance; vertical pressure distribution is presumed to be parabolic. Brush tyre design can be split into 3 instances: pure side-slip, pure longitudinal slip, and combined slip problem.

\subsubsection{LuGre tyre model} is a physics-based dynamic tyre model, first implemented in 1995 by de Wit et al. The surfaces should be in touch through elastic bristles. It's shown in Khal-Fig16 \cite{canudas_de_wit_new_1995}. Several trials used LuGre model for estimating friction force or friction coefficient. De-Wit et al. \cite{canudas-de-wit_new_2003} used a single wheel dynamic system with a lumped LuGre friction model and presented a new highway shift parameter. Using measured wheel angular speed information, they intended an online observer for car longitudinal speed and highway situation parameters \cite{canudas-de-wit_observers_1999, canudas-de-wit_new_2003}. Alvarez et al. \cite{alvarez_dynamic_2005} also used the same strategy to develop an emergency braking-control-based driving tyre friction system. They also used a single-wheel dynamic model for the forces along with lumped LuGre theory. Using measured rotor angular speed information, the LuGre model's internal state (z), longitudinal and relative velocity ($v_r$) are predicted. Chen et al. used a bike template and proposed the following observer to assess LuGre model's internal states \cite{chen_adaptive_2011}. In another research, Alvarez et al. \cite{alvarez_dynamic_2005} created a LuGre model-based adaptive friction prediction algorithm. They used a quarter car model in which forces were simulated using LuGre modeling and measured LuGre's vehicle speed and inner parameters using sliding mode observer. Measured states were angular wheel speed and longitudinal speed. Matusko et al. \cite{Khal_133} used LuGre model to clarify friction force dynamics. They used a single vibrant model with LuGre tyre model to assess the friction force. A Neural Network (NN) algorithm is also used to offset tyre friction design uncertainties. They used Lyapanov Direct Method to adjust NN algorithm parameters. In some research, the LuGre tyre system was used to create new dynamic models. Cleays et al. \cite{Khal_61} created a LuGre-based tyre models describing the longitudinal and lateral forces and alignment moment with a set of first-order differential equations appropriate for traction and ABS driving systems. The average deflection of the bristles in the lumped LuGre model (which is presented by $z$) is expressed as \cite{canudas_de_wit_new_1995}

\begin{equation}
\begin{array}{l}{\frac{\mathrm{d} z}{\mathrm{d} t}=v-\frac{|v|}{g(v)} z} \\ {F=\sigma_{0} z+\sigma_{1} \frac{\mathrm{d} z}{\mathrm{d} t}+\sigma_{2} v} \\ {\sigma_{0} g(v)=F_{\mathrm{c}}+\left(F_{\mathrm{s}}-F_{\mathrm{c}}\right) \mathrm{e}^{-\left(\frac{v}{v_{s}}\right)^{2}}}\end{array}
\end{equation}

where $v$ is the relative velocity between the two surfaces, $v_s$ is Stribeck velocity, $F_c$ is the Coulomb friction level, $F_s$ is the level of stiction force, $\sigma_0$ is rubber stiffness, $\sigma_1$ is rubber damping coefficient, and
$\sigma_2$ is the viscous relative damping.

\subsection{Prediction \& Estimation Models} 

Estimation of vehicle-dynamic variables is essential for safety enhancement, in particular for braking and trajectory-control systems. Active security systems such as anti-lock braking systems and electronic stability programs can be improved if the dynamic potential of a car is well known. For example, information on tire–road friction means a better definition of potential trajectories, and therefore a better management of vehicle controls. However, for both technical and economic reasons some fundamental data relating to vehicle-dynamics are not measurable in a standard car. As a consequence, dynamic variables such as tire forces and sideslip angle must be observed or estimated.

\subsubsection{Model Predictive Control}
The Model Predictive Control (MPC) has been examined for several implementations of vehicle control including engine \cite{di_cairano_mpc_2008, Hrovat_6, Hrovat_9}, transmission \cite{amari_unified_2008, Hrovat_11} and emissions \cite{schallock_model_2009}, mechatronic actuators \cite{Hrovat_14}, steering \cite{falcone_predictive_2007, di_cairano_steering_2010}, suspensions \cite{mehra_active_1997, giorgetti_hybrid_2006-1}, energy management \cite{Hrovat_19, Hrovat_20}, and thermal management \cite{vermillion_predictive_2011}. Several businesses, including Ford, BMW, Honda, Honeywell, PSA, Toyota, donated to and endorsed MPC studies. Since the early 1990s control techniques centered on MPCs have been studied at the Ford Motor Company on the inspiration of Professor Morari and his colleagues' landmark survey \cite{garcia_model_1989}. Some of the outcomes of these early attempts were subsequently released in \cite{Hrovat_6} under the Idle Speed Control, where MPC's benefits, as the capacity to manage actuator restrictions and load preview, were shown. Over the next few years a number of further exploratory studies including traction control \cite{borrelli_mpc/hybrid_2006}, semi-active suspensions \cite{giorgetti_hybrid_2006-1}, and Direct Injection Stratified Charge (DISC) engine \cite{giorgetti_hybrid_2006} were created. Although these research have largely been restricted to simulation research, important insights have been acquired into benchmarking MPC capacities in the car setting, in some instances leading to analytical breakthroughs, such as offering specific solutions for ideal semi-active suspension regulation \cite{giorgetti_hybrid_2006-1}. During the last decade, in completely drivable prototype cars at Ford, several MPC apps have been created, with controller requirements sometimes very comparable to those in manufacturing cars. Examples include traction control \cite{borrelli_mpc/hybrid_2006}, autonomous vehicle and stability control \cite{Hrovat_4, falcone_predictive_2007, falcone_mpc-based_2008}, idle speed control \cite{di_cairano_mpc_2008, di_cairano_model_2011}, and series HEV energy management \cite{di_cairano_engine_2011}. The active chassis with embedded lateral stabilization control can usually be summed up as a multivariable / multi-objective physical constraint problem. Model predictive control (MPC) in vehicle control systems \cite{trimboli_model_2009} has been more and more debated, especially in the area of car lateral stability controls because of their capacity to clearly deal with multi-objective optimisation and limitations. The computational effectiveness of the MPC for the active chassis integrated lateral stability control, particularly those running on resource-constrained embedded with computational resources has always been a major issue \cite{Hrovat_14}. While these MPC controllers still needed important computing funds, multiparametric programming \cite{bemporad_explicit_2002} combined with appropriate design steps \cite{di_cairano_mpc_2008, di_cairano_model_2010} led to feasible computational resources within automotive microcontrollers. 

MPC framework can be used to reflect efficiency degradation and mitigate the evolving degradation impacts at the lower levels of a system architecture. As mentioned in \cite{houjie_jiang_obstacle_2016}, the CMPC give less computational time than Non CMPC. It is noted that it is the first occasion that deep Time Delay Neural Networks (TDNN) have been used to simulate MPC behavior in the framework of constructing control algorithm as discussed in \cite{drgona_approximate_2018}. MPC \cite{hongbo_lateral_2017} has been created to incorporate optimum control efficiency with feedback control robustness. It also chooses activities by optimizing the cost function and taking into account the system dynamics and limitations similar to an optimal control. The increase in the speed and memory of the processor and development of new algorithms has made it possible to apply MPC to automotive control problems in recent years. MPC has several appealing characteristics for such apps. First, MPC enables multivariate feedback controllers to be designed with the same procedure complexity as single varying controllers. MPC also enables the definition of constraints on system inputs, states, and outputs during the layout stage that are then implemented by the controller. In addition, MPC enables for the definition of a controller-optimized objective function. Other advantageous characteristics of MPC include the ability to cope with time delays  \cite{todorovic_managing_2017}, the rejection of measured and recorded disturbances \cite{shi_algorithm_2017} and the use of prospective information \cite{pham_robust_2017}. Finally, MPC has a philosophical attraction since it embodies both optimization (receding horizon) and input adjustments, imitating numerous procedures naturally operating in this manner. 

MPC was applied for real-time path-following in \cite{falcone_predictive_2007, falcone_mpc-based_2008} and was used for Ts=50ms in a prototype car with steering actuation. Based on the present steering input and plant model, the autonomous vehicle from MPC has demonstrated to be more stable than the specifically designed steering robot for variable road surface friction \cite{falcone_predictive_2007}. In \cite{falcone_mpc-based_2008} the autonomous vehicle MPC controller was expanded to control steering and braking. The MPC methods introduced in tests include models of variable complexity (from two-track models with 4-braking torque control to single-track models with Yaw moment control \cite{falcone_predictive_2007}, \cite{falcone_mpc-based_2008}). Both nonlinear (NLMPC) model predictive controls as well as linear predictive control (LTVMPC) models have been created and tested on a dual lane shift maneuver with each model. While the NLMPC (1) uses a fully non-linear model with the Pacejka tire model, LTVMPC linearizes the model with each sampling time (7) and defines the cornering stifness in real time. These trials have shown that, although working in a broad spectrum of tire-road features, the proposed algorithm i) co-ordinates the use of steering and braking in presence of constrains, (ii) stabilises the vehicle regardless of road uncertainties, (iii) reproduces complicated counter-steering behavior carried out by qualified riders. MPC has demonstrated significant potential for utilization in automotive apps. However, some basic difficulties also need to be faced. First, it can be complicated to calibrate MPC, similar to other model-based methods. There are some measures to decrease the calibration effort (see for example \cite{di_cairano_model_2010}, \cite{Hrovat_42} and their references). Also, the MPC controllers are still too complicated for several apps with regard to computational constrains and low complexity explicit laws \cite{jones_approximate_2009, canale_set_2009}, or fast-approximated optimization algorithms \cite{wang_fast_2010} are essential. In addition, it is still difficult to ensure the sustainability of MPC a-priori, without unnecessary growth of algorithm complexity \cite{caruntu_predictive_2011}. Finally, several apps have important non-linearities, complicating the design and execution of MPCs.

\subsubsection{Kalman Filter} 
Kalman Filter is an estimator for what is called the linear-quadratic problem, which is the problem of estimating the instantaneous state of a linear dynamic system perturbed by white noise - by using measurements linearly related to the state but corrupted by white noise. The resulting estimator is statistically optimal with respect to any quadratic function of estimation error. An extended non-linear version of Kalman Filter is later introduced under the name Extended Kalman Filter (EKF). An EKF that takes a stochastic approach for the solution of the estimation problem is a popular candidate for accurate estimation results of vehicle velocities as it utilises available model information and high accuracy onboard sensor measurements \cite{li_variable_2011, baffet_estimation_2009}. However, EKF has its own shortcomings to overcome in this specific application. The states change rapidly during the vehicle running, and the latency of EKF should satisfy the requirement of control loop time of ABS and ESP that is usually chosen for 20 ms \cite{imsland_nonlinear_2007}. Moreover, EKF includes highly mathematical complexity and involves a large volume of data, because there are many matrix multiplication and inversion operations performed in each iteration. The excessive computational requirements for high order vehicle dynamics system become significant constraints and sometimes limit the use of EKF in automotive system \cite{antoniou_exploratory_2013}. To overcome the computational burden of EKF scheme, sliding mode observer (SMO) and non-linear state observer are proposed to solve the specific problem. Sliding Mode Observer (SMO) has been widely employed for vehicle velocities estimation because of its effectiveness in applications for computational efficiency as well as its robustness on parametric variations and modelling uncertainties \cite{stephant_evaluation_2007} and is discussed next.

\subsubsection{Linear, Non-linear, and Sliding Mode Observers} 

Braking and control systems must be able to stabilize the car during cornering. When subject to transversal forces, such as when cornering, or in the presence of a camber angle, tire torsional flexibility produces an aligning torque which modifies the original wheel direction. The difference is characterized by an angle known as ”sideslip angle”. This is a significant signal in determining the stability of the vehicle \cite{Joanny_Bulteau}, and it is the main transversal force variable. Measuring sideslip angle would represent a disproportionate cost in the case of an ordinary car, and it must therefore be observed or estimated. For example, \cite{nielsen_automotive_2000} presents linear and nonlinear observers using a bicycle model. \cite{venhovens_vehicle_1999} uses a Kalman filter for a linear vehicle model. \cite{stephant_vehicle_2003} presents a comparison of several linear and nonlinear observers. Performances obtained through simulation and experimentation using several classical pure lateral dynamics tests are presented, and the nonlinear observability problem and its relation to vehicle speed and cornering stiffness are discussed. 

The non-linear state observer for vehicle velocities estimation with input-to-state stability (ISS) \cite{Cao_14} guaranteed along with yaw rate is presented in \cite{grip_nonlinear_2008}. It is based on longitudinal and lateral accelerations and yaw rate measurements in addition to wheel angular speed and steering wheel angle measurements. Moreover, the sideslip angle is obtained according to the longitudinal and lateral velocities. However, the observer gains must be determined at each sample time. So the real-time performance of the estimation algorithm may be affected, and the choice of the observer gains may be interfered by the measurement errors. Additionally, a significant weakness of the observer is that the choice of observer gain mainly rely on the designer’s experience and it would be tuned for many times to obtain good estimation performance. In order to improve the real-time performance of the nonlinear state observer, a reduced-order observer for clutch torque estimation is discussed in \cite{gao_observer-based_2012}. The implementation of the designed observer benefits from its reduced order and also from that the observer gains can be chosen to be time-invariant \cite{Cao_17}.

\section{Advanced Vehicle Systems Technologies}

In advanced vehicle systems, some technologies are required in order to sustain a smooth flow of information between different nodes inside and outside the vehicle gathering data on the driver, environment, vehicle states, etc. These technologies are classified into three; Data-driven technologies where huge data is being processed and trained to be able to predict specific futuristic outputs based on present experiences, Communicational technologies which enable safety services and utilize various exchange of information between road travelers, and Sensing technologies that make use of physical properties in order to give some sensory information that can be fused or exchanged or used for decision making.

\subsection{Data-Driven Technologies}
\subsubsection{Statistical Approaches} weighted, multivariate and most up-to-date data mining engines \cite{Faouzi_6}. The most simple method for information combinations is the arithmetic mean approach among statistical techniques. This approach is not appropriate when information can not be exchanged or if the performance of estimators / classifiers is unlike \cite{Faouzi_6, hashem_optimal_1997, el_faouzi_heterogeneous_1997}.

\subsubsection{Probabilistic approaches} For example, Bayesian approaches are widely used for multisensor data fusion with Bayesian network and space models \cite{Faouzi_10}, maximum likelihood method and Kalman filtered DF \cite{huang_adaptive_2010, Faouzi_12}, theory of options \cite{dubois_possibility_1988}, evidence reasoning and specifically evidence-based theory \cite{dempster_upper_1967, dempster_generalization_1968, shafer_mathematical_1976}. This latter technique can be considered as a widespread Bayesian approach \cite{dempster_generalization_1968, shafer_mathematical_1976, el_faouzi_improving_2009}.

\subsubsection{Artificial intelligence} Artificial cognition and neural networks including artificial intelligence, genetic or neural algorithms. The latter approach serves as a tool for the collection of classifiers or estimators as well as the combining framework of classifiers / estimators in many applications \cite{Faouzi_6, hashem_optimal_1997}. 
\paragraph{Supervised Learning}
Supervised learning is based on a model which can predict an output for new input data. It is based on prior knowledge of the results of certain inputs. Take, for instance, a doctor who wants to forecast if a patient has cancer and if so whether it is benign or malignant. He may analyze the history of the patient and compare it with others that go far into the family's history of health. In the end, a model can be created that provides a reasonable but uncertain outcome, predicting the possible outcome. 

In \cite{Moujahid_12}, there is a mathematical definition: given the number of inputs usually called training set $X$, we have a corresponding function $f$ value. The aim is to find the best hypothesis for supervised learning, which allows us to find the close value of $f$ on a given $X$. Supervised learning techniques always involve works of classification and regression, only the input differentiates between the two: discrete for classification while continuous for regression. The first consists of a defined and finite number of values between the two, while the latter consists of a range. In this ML category, the work of Patrick Tchankue et al. \cite{Moujahid_12}, Claire D’Agostino et al \cite{Moujahid_13, dagostino_learning-based_2015}, Yi Hou et al. \cite{yi_hou_modeling_2014} and Brendan Morris et al. \cite{Moujahid_26} fit.

\paragraph{Unsupervised Learning}
The lack of accurate knowledge of what the data contain or what the end purpose is, characterizes uncontrolled learning. This is also the main difference with supervised learning, in which a known response can be predicted. It is not possible to assess its result due to the absence of fixed labels. Unsupervised learning will teach the machine how to collect data and to try to explain what is therein, even if they are structured or contained in a manner that is not clear. 

The training set $X$ has no known output value (function $f$) from a mathematical point of view \cite{noauthor_machine_2018} and its ultimate purpose is to try and create one. An example of what unsupervised learning can do is describe what lights, vehicles, people etc. are at crossroads. You can also try to split and group various types of vehicles: bicycles, coaches, buses. What's most significant is that you do not know what the crossroads contain in advance; you just see that all vehicles have different characteristics. They move at different speeds, for example, and just when you are on the road you decide to call them vehicles without really knowing what a vehicle is, because it's just one possible classification method. Because of all these aspects, we talk about extracting features when applying an unattended algorithm as unsupervised learning: unseen patterns or labels are found, common features are extracted from individuals being considered. Because of the ability to move, both vehicles and people are characterized by a distinguishing characteristic from traffic lights; it can also be featured in two distinct clusters \cite{noauthor_unsupervised_2017}.

\paragraph{Reinforcement Learning}
Reinforcement learning is a specific field of machine learning based on actions and numerical rewards for the achievement of a specific goal \cite{sutton_reinforcement_2018}. The important point is that the person taking action, named agent, does not know what actions are good or bad in a certain governing world, named the environment, but she knows which ones give the highest rewards in the test. 

Furthermore, a policy $p$ is defined as the way to a solution. Mathematical problems are always modeled on Reinforcement-learning as part of Markov Decision Process (MDP) \cite{sutton_reinforcement_2018}. MDP provides the mathematical rules for decision-making issues, both for their description and solutions. Consider the parking action: this could be thought of as an MDP. The environment would be the car park; the agent would be the wheel system and the reward is connected with the correct positioning of the car, the state. In fact, reinforcement learning is mainly used for parking assistance, but it would also be used for evasive steering actions in the future.

\paragraph{Deep Learning}
Deep learning is a multi-layered approach to artificial intelligence (AI) with a final profile-chart \cite{Moujahid_27}. Deep learning techniques are generally used with the aim of creating and representing the model by the artificial neural network (ANN)  An ANN is always made up of an input layer and an output layer. The deep learning algorithm is used between several intermediate layers. Deep learning can be used to recognize speech, classify and detect objects, recognize designs, robotics and self-driving automotives. Although Deep Learning requires enormous training data and computational capacity, it's still one of the best options for real-time learning \cite{krizhevsky_imagenet_2017}. 

There are three main models in deep learning which are Recurrent Neural Networks (RNN), Convolutional Neural Networks (CNN) and Deep Belief Networks (DBN) \cite{Moujahid_29}. Deep learning plays an important role in the world of image processing. Clearly analysing the key parts of an image is possible with the ability to represent images using multiple hidden-level neural networks. This is why the driving alertness monitoring system is based on in-depth learning techniques that point to the driver's face using cameras. Deep learning is used as well as on the basis of computer vision models for obstacles, road paths and pedestrian detection \cite{Moujahid_30, jia_obstacle_2016, Moujahid_32, lee_real-time_2016}.

\subsection{Communicational Technologies and Architectures}
Data are valuable beyond the confines of a single vehicle, so intelligent transportation systems need connectivity that works at high velocities, long range, and with dynamic peers. Intervehicle networks share data among vehicles and  infrastructure, facilitating data collection and optimization at scale. Improvements in network scalability, routing efficiency, data security and quality of service have made wireless networking tenable, allowing for the use of mesh and topdown networking approaches for data’s movement from within vehicles to remote computing devices.

\subsubsection{Mesh Networks}
Vehicular mesh technologies support the needs of transportation data sharing, connecting vehicles and infrastructure in transient, ad-hoc neighborhoods. This section considers mesh networks’ enabling communication standards and data routing protocols. Mesh networks can be classified into communication standards and broadcast types.
	
Standardization is a critical enabler of connected vehicles. A leading standard supporting traffic safety and efficiency is Dedicated Short Range Communication (DSRC). Several studies compare different standards commonly used in the United States and Europe~\cite{Siegel_36, Siegel_37, bai_towards_2012, Siegel_39, faezipour_progress_2012, faezipour_progress_2012, Siegel_40, Siegel_41, Siegel_42, Siegel_43, Siegel_44, festag_standards_2015, Siegel_46, Siegel_47, festag_standards_2015, Siegel_48}.
	
Connected messages have varied sensitivity to timeliness, data protection, and network range. It is imperative to choose an appropriate broadcast protocol to assure application performance. For example, the Wave Short Message Protocol (WSMP) enables the use of smaller packets for time-sensitive safety and convenience applications \cite{bai_towards_2012}. In vehicle-to-vehicle or vehicle-to-infrastructure applications, messages may broadcast openly, or receivers may subscribe to specific topics. In this case, publishers push event data to a network without a target, and recipients accept select message types \cite{Siegel_49}. Various works have investigated different broadcast types and their usability in vehicle-to-vehicle and vehicle-to-infrastructure communications \cite{papadimitratos_vehicular_2009, casteigts_communication_2011, Siegel_36, papadimitratos_vehicular_2009, casteigts_communication_2011, papadimitratos_vehicular_2009, Siegel_51, biswas_vehicle--vehicle_2006, Siegel_53,Siegel_53, Siegel_54}.
	
	\subsubsection{Cellular Networks}
	Widespread Machine-to-Machine (M2M) connectivity has commoditized cellular bandwidth and hardware \cite{Siegel_55}, lowering costs and allowing vehicles to connect to one another and to remote data sources and sinks directly and indirectly. Such direct connectivity uses a vehicle’s integrated modem to stream data to a remote server. Unlike mesh and short-range networking, direct Vehicle-to-Internet cellular connectivity is robust and capable of sharing data when traffic density is sparse \cite{Siegel_56, wilhelm_cloudthink:_2015}. Different vehicle communication technologies are compared in \cite{mumtaz_cognitive_2015, mukherjee_integrating_2015, Siegel_59, inam_feasibility_2016, wang_cellular_2014, Siegel_62, Siegel_56, wilhelm_cloudthink:_2015, faezipour_progress_2012}.

\section{Autonomous Driving Modules}
Nowadays, the automotive industry is moving forward with the term of advanced vehicles into self-driving vehicles, requiring a lot of computational resources and integration of automation modules. These modules are mainly classified into a hierarchy of three stages where each stage is responsible for processing information to the next stage. The perception and data fusion modules are utilized for processing raw sensor data into valuable data for the planning module. These raw sensor data range from random moving vehicle states to collection of images from different cameras or radars, under the term Computer Vision. This module processes the raw sensor data from the stage of data collection to preprocessing stage that provides a set of features for each sensor, to detection and classification, and a combination of these data. The module then passes information to the Planning module. 

The planning module's first route planning is done to select the optimum path with minimum cost. Based on the generated optimum path, behavioral decision planning takes place where the most proper conduct of the vehicle following the path is planned to ensure safe tracking and consider other road travelers’ behavior. The planned path is later fed to a low-level feedback controller to ensure the planning is executed. After a behavioral decision is planned, ensuring no collision or false lane change, motion planning is carried out to generate a complete trajectory of the ego vehicle.

\subsection{Perception and Data Fusion}

In the perception context, the process could be analyzed from the perspective of three low-medium-high levels of perception and data fusion. In low-level applications, new raw information is created from different sources. One of the classical fusion applications in computer vision is stereo vision, where information from two cameras is combined, creating new information, i.e., the disparity map. In \cite{Garcia_1, Garcia_2}, stereovision is used to perform pedestrian detection, and in \cite{Garcia_3}, this information is used together with a laser scanner for this purpose. Other low-level approaches provide sensors to enhance the global position systems (GPS) \cite{Garcia_4}.  

In medium-level approaches, the first preprocessing stage provides a set of features for each sensor. These sets are combined to perform obstacle detection and classification. In \cite{Garcia_5, Garcia_6}, authors present different approaches, whether combining or not, the different features of the different sensors comparing results. In high-level fusion approaches, detection and classification are performed for each sensor independently, and the final stage combines the detections according to the accuracy of the detections and sensors. In \cite{Garcia_7}, Adaboost vision is used for pedestrian detection and Gaussian Mixture Model Classifier (GMMC) for laser scanner-based pedestrian detections. Existence probability based on Dempster-Shafer is used in \cite{Garcia_8, Garcia_9}. Other approaches among Intelligent Vehicles researches
use data from a laser scanner to detect regions of interest (ROI) in images and computer vision to classify among different obstacles included in these ROIs. In \cite{Garcia_10}, raw images with the SVM machine learning method are utilized. Authors in \cite{Garcia_11} use Histogram of Oriented Gradients (HOG) features; moreover, other sensor fusion approaches take advantage of the properties of various sensors differently. In \cite{Garcia_13}, information from a stereovision camera and a laser scanner is combined. First, the application uses stereovision information to locate the road. Then it uses this information to remove those irrelevant obstacles for the application (i.e., outside the road). Next, it constructs a set of obstacles using the information from both sensors, and tracking is performed using a Kalman Filter approach. In \cite{Garcia_14}, it is used information from the laser scanner to search particular zones of the environment where pedestrians could be located and visibility is reduced, such as the space between two vehicles, and performs detections using the visual approach. Finally, in \cite{Garcia_15}, a laser scanner and radar approach are used for obstacle detection and tracking, and a camera is used to show the results. Obstacle classification differentiates between moving and non-moving obstacles by computing Mahalanobis distance among the clusters given by the laser scanner.

Computer vision is a combination of both radio detection and ranging and light detection and ranging, which is at the bleeding edge of innovations that empower the advancement of advanced driver assistant systems. Advanced driver assistant systems are intended to expand drivers' situational mindfulness and street well-being by giving fundamental data, cautioning, and programmed intercession to decrease the likelihood or seriousness of a mischance. Contingent upon the well-being perspectives, ADAS can be founded on sensor frameworks, for example, radio detection and ranging, mono camera, light detection, and ranging-based frameworks. 

Presently camera-based ADAS improvement is a standout amongst the effectively investigated region. Camera-based ADAS may be intended to utilize one or mono as well as more, i.e., mono and stereo-based camera sensors.

Some examples of utilizing data fusion in the perception process are stated below:
\subsubsection{Vehicle Detection}
\cite{chong_integrated_2013} proposed a method to detect and track vehicles in urban areas. This paper introduces a novel algorithm for the ongoing operation of vision-based vehicle recognition frameworks. This algorithm includes two principle segments: detection of vehicles and also tracking of vehicles. Vehicle recognition is accomplished by utilizing vehicle shadow highlights to characterize a region of interest (ROI).

\subsubsection{Pedestrian Detection}
\cite{dixit_pedestrian_2015} proposed a system to detect pedestrians using Haar cascade, frontal face detection algorithm, and Friendly ARM board S3C2440. Figure 6 shows a workflow of the proposed system. The preprocessing method includes the removal of unwanted noises or distortions from input data. Feature extraction is to identify the targeted object.

\subsubsection{Traffic Sign Detection}
\cite{Karth_12} proposed a system to detect pedestrians using laser scanners and data fusion techniques. Sensors are an important component of advanced or automatic driver assistant systems. The absence of dependable sensors ready to satisfy the prerequisites for street well-being algorithms and applications prompts the need to consolidate them to make a combination technique ready to defeat the restrictions of a solitary arrangement of sensors.

\cite{yanjun_fan_traffic_2015} proposed a traffic sign recognition and classification system. Traffic signs incorporate numerous valuable natural data which can enable drivers to discover the difference between the street ahead and the driving necessities. Accordingly, an ever-increasing number of researchers have focused on the issues of acknowledgment of the activity signs by utilizing computer vision and machine learning methods. Also, the traffic signs detection algorithm has become a vital piece of Advanced Driver Assistance Systems (ADAS). A novel traffic sign detection algorithm, in light of machine vision and machine learning procedures, is proposed in this paper.

\subsubsection{Driver Monitoring System}
A driver monitoring algorithm for the ADAS system was proposed in \cite{Karth_17}. This implementation has a few phases. The input image is captured from the camera and converted into grayscale because the grayscale image is well suited for face and iris detection. The face is detected by using Haar feature selection and Adaboost training. Grayscale human image has some unique characteristics, such as the eye region being darker than others and the nose region being brighter than others.

\subsubsection{Object/Obstacle Detection}
An example of an object detector system on an embedded board was proposed in \cite{lee_optimization_2016}. This technique intends to accomplish organized enhancement of street object detectors by preparing memory and time of convolutional neural network models executed on an installed load-up. A question finder is created called the single-shot multi-box identifier that contains a VGGNet-based framework and a multi-box identifier using various scale incorporate maps.

\subsubsection{Lane Detection}
Lane detection and departure warning system are critical modules in the ADAS system. Real-time lane detection and departure warning system on an embedded platform was proposed in \cite{lee_real-time_2016-1}. This proposed method uses Inverse Perspective Mapping (IPM) and the Kalman filter. Input images captured from the camera are transformed into top-view or bird-eye images using the Inverse perspective mapping method. This method will remove the effect of perspective in an image.


\subsection{Planning} 

\subsubsection{Route Planning}
The decision-making scheme of a car must choose a path from its present location through the highway network to the desired location at the highest level of decision-making. If the highway network is presented as a guided graph with borders equal to the cost of passing a highway section, the issue of identifying a minimum-cost path on a road network graph can be defined. However, the road network graphs can include millions of edges that do not include the usual shortest route algorithms, such as Dijkstra \cite{dijkstra_note_1959} or A* \cite{Paden_33}. The issue of practical route planning has drawn significant attention in the transportation scientific community, leading to the development of a family of algorithms that deliver an ideal path in milliseconds to a continent-scale network after a one-time preprocessing stage \cite{Paden_34, geisberger_exact_2012}. For an extensive investigation and comparison of practical algorithms that can be used to schedule paths for both human-driven and self-driving cars effectively, see \cite{delling_route_2018}.

\subsubsection{Behavioral Decision Making}
Driving manuals require qualitative behavior in particular driving environments. The independent advanced vehicle must be prepared to travel on the chosen path after establishing a path scheme and communicating with other traffic respondents under highway regulations and riding conventions. Given the series of highway sections specifying the path chosen, the cognitive layer selects the proper driving conduct at every moment, depending on the conduct of other traffic respondents, street circumstances, and infrastructure signals. For example, the behavioral layer commands the vehicle, when reaching the stop line before an intersection, to stop, to observe other vehicles, bikes, and pedestrians ' behaviors at the intersection, and let the vehicle go once its turn has arrived. 

As riding environments and attitudes are represented as finite sets, a natural strategy to automate this choice is modeling each conduct as a state in a final state machine with apparent driving context changes such as relative position in the scheduled path and neighboring cars. In reality, the DARPA Urban Challenge embraced finite state-based systems combined with various heuristics that regarded driving situations as a system for behavioral command by the majority of teams \cite{Bast_9}. 

However, real-world driving, particularly in metropolitan areas, is defined by uncertainty about the intentions of other road users. The issue of intention to predict and estimate the potential trajectories of other cars, motorcycles and cyclists has also been examined. Machine learning-based methods, e.g. gaussian blend models \cite{bauer_search-space_2016}, Gaussian process regression \cite{Bast_38}, and model-based approaches allegedly used in the self-driving intention prediction scheme of Google \cite{bauer_sharc:_2009} and model-oriented solutions to straight estimating signal measurement targets \cite{bauer_combining_2010, bauer_experimental_2011} are among the solutions suggested. This uncertainty in the conduct of other traffic respondents is generally considered in decision-making using probabilistic planning formalisms, such as the Markov Decision Processes (MDPs) and their generalization. For instance, \cite{Bast_42} defines the issue of cognitive decision-making within the MDP structure. 

Several works \cite{Bast_43, baum_energy-optimal_2013, Bast_45, Bast_46} fully use a Partially Observable Markov Decision Process (POMDP) to model unconsidered driving and pedestrian intentions and suggest particular estimated response strategies.

\subsubsection{Motion Planning}
When determining which behavior, such as cruise-in-lane, change-lane, or turn-right, the behavior selected should be transformed into a path or trajectory followed by the low-level feedback control system. The motion planning system's task is to discover this path or trajectory. The path or trajectory that results must be easy for the rider, dynamically feasible for the vehicle, and avoid a collision with obstacles detected by onboard sensors. 

The motion planning task of an autonomous vehicle represents the question of standard motion planning, as stated in the robotics literature. Exact solutions to the motion planning problem are computationally intractable in most circumstances. Thus, numerical approximation methods are generally used in practice. Variational methods are one of the most popular numerical methods as differential regression in the functional space. Graph-search methods are designed to create graphic discretization of the state space and search for the shortest path with graph-search methods and incremental tree-based methods that progressively construct the tree of a reachable state from the initial state and then select the best branch of such a tree.

\subsection{Decision Making}
Driverless cars are autonomous decision-making systems that process a stream of observations from onboard sensors such as radars, LIDARs, cameras, GPS/INS units, and odometry. Together with prior knowledge about the road network, road rules, vehicle dynamics, and sensor models, these observations are used to automatically select values for controlled variables governing the vehicle’s motion. Intelligent vehicle research aims to automate the driving task as much as possible. The commonly adopted approach to this problem is partitioning and organizing perception and decision-making tasks into a hierarchical structure. In this section, we describe the decision-making architecture of a typical self-driving car and comment on the responsibilities of each component.

\subsection{Vehicle Control}
In order to execute the reference path or trajectory from the motion planning system, a feedback controller is used to select appropriate actuator inputs to carry out the planned motion and correct tracking errors. The tracking errors generated during the execution of a planned motion are due in part to the inaccuracies of the vehicle model. Thus, a great deal of emphasis is placed on the robustness and stability of the closed-loop system. Many effective feedback controllers have been proposed for executing the reference motions provided by the motion planning system. 



\section{Open Issues, Challenges and Research directions}
\subsection{Open issues and challenges}
Challenges in the advanced vehicle safety domain remain very broad, ranging from accuracy and efficiency of objects detection, management of sensory data, appropriate choice of decision-making techniques, validation and testing of software components in addition to reliability and quality of such components, to network performance and data handling in a vehicular communication context. Moreover, some open issues in the advanced vehicle safety research domain rely upon investigating the correspondent network bandwidth of each utilized protocol, side-slip angle, and tire-road friction coefficient estimation. 

In the object detection field of advanced vehicle safety, the context is changed based on the type of the object, which could be a vehicle, a road lane, a pedestrian, or a collision, given that the latter is more complex than previous types. Employing fast road speeds to allow increased traffic flow will burden such techniques and require fast and critical algorithms for timely decision-making that cope with such a dynamic environment. Challenges of managing sensory data rely mainly on the need for real-time processing capabilities and the fact that deep learning algorithms are storage and compute-intensive. Authenticity and data redundancy remain challenging, expanding from hardware to communication-context limitations. Such challenges regarding data management could be approached in many ways:
\begin{itemize}
\item increasing computational and communication resources,
\item utilizing crowd-sourcing and crowd-sensing techniques,
\item sharing sensory data across different nodes
going under the trade-off bargain between the number of used sensors and the efficiency of data processing
\end{itemize}

Existing decision-making mechanisms in autonomous cars can be categorized into machine learning, deep learning, artificial intelligence, multi-policy decision-making, and the Markov decision process. 

The main challenge occurs when choosing the most appropriate technique. Moreover, the unpredictability of the surrounding environment and human behavior burden such choices in addition to the difficulties of detecting faults and malfunctions of said systems and making optimal decisions in this regard. Context- and situational-aware objection detection and decision-making algorithms could be utilized as a workaround to the previously stated challenges. The software components context in advanced vehicle safety has a variety of challenges regarding validation, testing, safety, reliability, and quality. 

These challenges could be summarized as the need for complete requirements that allow developers to test-check their components in addition to the complexity of such operations given their dynamic and non-deterministic nature.
Regarding safety and reliability, legislation and validation cycles are still vague and ambiguous on the deployment level. At the same time, substantial budget requirements and several unforeseen scenarios remain a great challenge and a barrier in the face of software quality. Approaching such challenges in the software context could be done through functional division of software/hardware components, limited operational concepts, and machine learning methods that would allow a better realization of validation and testing for such components. More approaches would include
\begin{itemize}
\item defining short-term safe missions,
\item developing sophisticated algorithms for small missions,
\item deciding fail-safe rather than anticipating unpredictable outcomes, and
\item defining goal-oriented software quality testing.
\end{itemize}
In time-critical connected vehicle applications such as safety or vehicle control, messaging must be low latency and delay-bounded, requiring complex network architectures that utilize up-to-date routing information. Such real-time changing network conditions further challenge decentralized information flow concerning data transmission and handling. Processing of such data even with accuracy constraints, processing such data remains a significant challenge in vehicular communication that requires specialized handling tools. 

Pre-processing metrics may make data useful in a compact form, reducing the size of digitally duplicating vehicles. These metrics may be aggregated from multiple vehicles to minimize storage and analysis cost and complexity. These architectures strike different balances of power, computation, data accuracy, and feasibility. For tire-road friction coefficient estimation, specifically side-slip angle, challenges remain in the measurement due to the costly and noisy sensors used, which leads to estimation by modeling. In order to overcome such challenges, storing raw data could be feasible but remain costly; map reduction techniques would then be necessary to allow scalable data analysis and pattern identification. 

In automated driving and connected vehicles, the range of estimation must be extended to predict the distance and relative speed of other vehicles to decide on the next position and dynamic states. More challenges include dealing with multi-sensor information redundancy, where estimation schemes should be able to differentiate between valuable and non-useful information in addition to using minimal sensor information. It is now reasonable to believe that extended state observers and adaptive observers estimation strategies will be very effective in solving the estimation problem with unknown inputs such as missing road information. The field of tire-road friction estimation may also be extended to include weather effects where data-driven-oriented methods, such as artificial neural networks (ANN) and fuzzy logic (FL), become helpful to be adapted. Such approaches require large training data sets that remain a significant challenge facing tire-road friction estimation in the research field. 

\subsection{Research directions}
Advanced vehicle safety research has grown significantly in the past decade, including different solutions and approaches to different problems and challenges. Examples of such approaches include many advances in the utilization of communication advantages in the vehicular context, such as employing hybridized networks, utilization of cellphones as probes, Safety-as-a-Service (SaaS) frameworks, Road Safety Index (RSI), and cooperative adaptive cruise control (CACC) in platoons. 

Moreover, with the rapid rise of machine learning techniques, employing such techniques in communication structure has been introduced as many approaches, such as heterogeneous wireless networks (HWN) and Fast Machine Learning in 5G frameworks. Soon, in communication design, hybridized networks will combine radio technologies to enable low-latency, and extended-range connectivity through parallel use and handoff mechanisms \cite{Hazem_66}. 

Other approaches may use computation to emulate “impulse” reactions in the car so that time-sensitive applications can operate despite high-latency connectivity \cite{Hazem_181}. Eventually, 5G cellular will support high bandwidth, long range, and low latency direct-to-cloud connectivity, minimizing requisite vehicle density and improving the richness of data stored in vehicle mirrors \cite{Hazem_182}. Feasibility assessments show promise in using this technology in real-time vehicle teleoperation or for safety services today using V2V technology \cite{Hazem_60, Hazem_182, Hazem_183}. Particular focus is placed on vehicle-based monitoring systems, such as driving behavior and style recognition, accident detection, and road condition monitoring systems using cell phones as probes for sensing and exchanging such information. In contrast, data security regarding such an approach is still an open issue \cite{Hazem_5}. 

Safe-aaS is one of the first attempts in its domain, where multiple end-users dynamically receive safety-related decisions. An end-user enjoys the pay-per-use service of Safe-aaS without concern about the back-end process. Safe-aaS is based on service-oriented architecture, where business entities such as vehicle owners, sensor owners, safety service providers, and end-users are involved. We introduce the term decision virtualization, which enables multiple end-users to access the customized decisions remotely \cite{Hazem_7}. 

Road traffic injuries (RTIs) and Road Safety Index (RSI) are realized as primary causes of public health problems at global, regional, and national levels. Therefore, the prediction of road traffic death rate will be helpful in its management. Based on this fact, we used an artificial neural network model optimized through a Genetic algorithm to predict mortality. In this study, a five-fold cross-validation procedure on a data set containing a total of 178 countries was used to verify the performance of the models. The best-fit model was selected according to the root mean square errors (RMSE). Genetic algorithms as powerful models that have not been introduced in the prediction of mortality in previous studies showed a considerably high performance \cite{Hazem_8}. Cooperative Adaptive Cruise Control (CACC) is a very intriguing and vast area of automotive applications where the most essential and vital aspects are reviewed in \cite{Hazem_9}. 

The most challenging issues for real-world deployment of CACC in terms of aspects such as communications, driver characteristics, and controls are identified and investigated. Employing machine learning techniques in communication services includes an intelligent 5G heterogeneous wireless network architecture, a reinforcement learning-based Dedicated Short Range Communication (DSRC) for Vehicle-to-Vehicle (V2V) and millimeter wave (mmWave) for Vehicle-to-Infrastructure (V2I) communications \cite{Hazem_13}. 

Another approach under the term Fast Machine Learning (FML) is also proposed as an online learning algorithm addressing the problem of beam selection with environmental awareness in mmWave vehicular systems. In particular, the problem is modeled as a contextual multi-armed bandit problem. Next, a lightweight context-aware online learning algorithm, namely fast machine learning (FML), with proven performance bound and guaranteed convergence, is also proposed. In the tire-road friction coefficient estimation problem, neural networks (NN) techniques are utilized outside the communication context. New estimation methods are being developed by taking advantage of layered neural networks for the side-slip angle estimation since it has been difficult to utilize a control system due to the high cost of special measurement devices needed \cite{Hazem_10,Hazem_11,Hazem_12}. 

In continuation of deploying data-driven techniques within decision-making components, a data-driven modeling approach to capture the lane change decision behavior of human drivers is proposed in \cite{Hazem_6}. Data is collected with a test vehicle in typical lane change situations and train classifiers to predict the instant of lane change initiation concerning the preferences of a particular driver. Furthermore, this decision logic is integrated into a model predictive control (MPC) framework to create a more personalized autonomous lane change experience that satisfies safety and comfort constraints. One can conclude that research directions aiming toward utilizing data-driven techniques while taking advantage of the massive capacity of communication protocols will be the most significant research efforts within advanced vehicle safety in the next decade.



\ifCLASSOPTIONcaptionsoff
  \newpage
\fi

\bibliographystyle{IEEEtran}
\bibliography{main}



\end{document}